\documentclass[12pt]{article}
\usepackage{fullpage,psfig,amsfonts,latexsym,amsmath,epsf,latexsym,amssymb}
 
\input{Qcircuit}

\newtheorem{definition}{Definition}
\newtheorem{theorem}{Theorem}

\newcommand{\R}[0]{{\mathbb{R}}}

\newcommand{\Z}[0]{{\mathbb{Z}}}

\def\C{{\mathbb{C}}}
\def\Z{{\mathbb{Z}}}
\def\R{{\mathbb{R}}}
\def\Q{{\mathbb{Q}}}
\def\N{{\mathbb{N}}}

\newcommand{\cU}[0]{{\mathcal U}}

\newcommand{\cH}{{\cal H}}

\title{{\bf 
On the Computational Power of\\ Molecular Heat Engines}}

\author{Dominik Janzing\thanks{email: janzing@ira.uka.de}\\[1ex] 
{\small IAKS Prof. Beth, Arbeitsgruppe Quantum Computing,} \\
{\small  Universit{\"a}t Karlsruhe,
Am Fasanengarten 5,} \\ {\small 76\,131 Karlsruhe, Germany}}

\begin{document}

\maketitle

\abstract{A heat engine is a machine which
uses the temperature
difference between a hot and a cold reservoir to extract
work. 
Here both reservoirs are quantum systems
and a heat engine is described by a unitary transformation
which {\it decreases} the average energy of the bipartite system. 
On the molecular scale, the ability of implementing
such a unitary heat engine is closely connected
to the ability of performing logical operations and classical 
computing.
This is shown by several examples:

(1) The most elementary heat engine 
is a SWAP-gate acting on $1$ hot and $1$ cold two-level systems 
with different
energy gaps. 

(2)  An optimal unitary heat engine 
on a pair of 3-level systems 
can directly implement
OR and NOT gates, as well as
 copy operations.  The ability to implement this
heat engine on each pair of 3-level systems taken from the 
hot and the cold ensemble therefore allows universal classical
computation. 

(3) Optimal heat engines operating on
one hot and one cold oscillator  mode with different frequencies
are able to calculate 
polynomials
and roots approximately.

(4)
An optimal heat engine acting on $1$ hot and $n$ 
cold $2$-level systems with different level spacings 
can even solve the NP-complete problem KNAPSACK.
Whereas it is already known that the determination of ground states
of interacting many-particle systems is NP-hard, the optimal 
heat engine
is a thermodynamic problem which is NP-hard even 
for $n$ {\it non-interacting}
spin systems. 
This result
suggests that there may be complexity-theoretic 
limitations on the efficiency of
molecular heat engines.}  

\section{Introduction}

One of the most important consequences of the second law of
thermodynamics is the statement that the heat energy of a bath with
{\it uniform} temperature cannot be converted into other forms of energy.
Instead, systems with different temperatures  are needed.
Machines using temperature differences between two or
several heat reservoirs are called {\it heat engines}. Here we
consider hypothetical heat engines on the quantum scale 
that ``extract'' energy from a collection of 
elementary quantum systems with different temperatures.
An appealing feature of thermodynamic machines on the
quantum scale is that their relation to
information processing devices become more obvious.
This is already seen in cooling algorithms which have been
proposed in the context of NMR quantum computing \cite{SchulVaz}, 
where
the analogy between {\it initialization of bits} and {\it cooling}
is apparent\footnote{This analogy is a basic insight for
the thermodynamics of computation \cite{Landauer:61} and Landauer's principle
saying that the erasure of of bit of information `wastes' 
the energy $\ln 2 \,\,kT$, where $k$ is Boltzmann's constant and
$T$ the reference temperature of the bath which absorbs the 
erased information.}. 
Another {\it Gedankenexperiment} which shows that
a memory of a computing device can play the role
of a thermodynamic reservoir is Szillard's hypothetical 
engine \cite{Szilard} which extracts work from
a reservoir with uniform temperature for the cost of 
writing on an initialized memory (the `cold' reservoir). 
From a modern point of view, it is 
natural to replace hot and cold reservoirs by quantum
registers. By identifying  qubits with physical 
two-level systems one obtains well-defined thermal
equilibrium states which allows the proper definition of `hot' and `cold' 
qubits. After having represented hold and cold reservoirs
by quantum registers, heat engines which transfer entropy from the
hot to the cold reservoir are {\it quantum operations}
and it is straightforward to ask for the complexity of
these operations compared to the complexity of {\it computing}
steps. This shall be our subject. As the article will show, 
one can construct many instances of molecular systems where 
heat engines are close to logical operations  and 
computing devices.

First we state more clearly what  we mean by ``extracting energy''
from a physical system. 
\begin{definition}{\bf (Energy Gain of a Unitary)}
\label{ExtrDef}
Given 
a quantum system with Hilbert space $\cH$ 
with Hamiltonian $H$, i.e., a self-adjoint operator
$H:\cH \rightarrow \cH$,
and 
the state be  a density operator $\rho :\cH \rightarrow \cH$. 
Then we say that a unitary 
$U$ extracts energy if and only if
\begin{equation}\label{Aver}
E_{gain}:= tr(U\rho U^\dagger H)-tr(\rho H)>0\,.
\end{equation}
\end{definition}

\noindent
Below we will consider ``unitary heat engines'' 
where $\rho$ is the product state
$\rho=\rho_A \otimes\rho_B$ of a bipartite system $A,B$ and 
$\rho_A$ and $\rho_B$ are thermal equilibrium states
with different temperatures $T_A$, $T_B$, respectively. 

Definition \ref{ExtrDef} requires some clarification. First of all one has to ask
where the energy goes. If it is 
transferred to the environment, the energy extraction 
will in general not perform any useful work. 
However, we assume that
it is transferred to some {\it target} system which is not explicitly  
included into the model. Consider, for instance a two level system
with energy gap $E$  
in the
 excited state $|1\rangle$. Transferring it to its lower level
$|0\rangle$ releases the energy $E$. 
If the process would be 
implemented by stimulated emission, for instance, 
the energy would be absorbed by the stimulating field mode.
We will not consider the problem how the released energy could really
perform some {\it useful} work, instead,  we 
we only demand that the unitary
process
$U$ in ineq.~(\ref{Aver}) {\it lowers} the average energy of the system. 
It should be emphasized that we only allow unitary transformations
here instead of  general completely positive (CP)
trace-preserving operations  \cite{Da78}, since it is commonly believed that all processes are unitary
provided that a sufficiently large environment is taken into account. 
By allowing general CP-operations one would therefore implicitly
allow {\it information} transfer between the 
system and its environment. The fact that the operation
on the {\it whole system} involves 
information processing would then be obscured by the restriction
to a {\it non-informationally} complete subsystem.  

One may be surprised why the target system, i.e., the energy sink
(like the field mode in the example above), 
does not explicitly occur 
in the description. From the
fundamental point of view 
one would expect a unitary operation on a system which includes
the target. 
The problem is that the thermodynamics in such a model
depends strongly on the assumptions on the physics of the target system.
If the latter 
acts not only as an energy sink but 
also as an {\it entropy} sink, the {\it free energy}, i.e.,
the thermodynamically valuable energy, is not necessarily increased
since the latter is a difference between energy and entropy multiplied
by the Boltzmann constant and a given reference temperature. 
The latter temperature determines therefore the gain of 
usable energy. 

In \cite{JWZGB} we have considered thermodynamic
models consisting of a hot and a cold system as well as a target system
which is driven to a non-equilibrium state by a energy conserving
transformation on the  whole {\it tripartite} system. The question, which
resources are sufficient to prepare
a desired 
non-equilibrium state in the target system has extensively been 
studied in \cite{JWZGB} in the context of a quasi-order of
thermodynamic resources. We will show in Section~\ref{Just} that
energy conserving transformations  on the tripartite system
can approximately lead to unitary transformations 
 by restriction to the hot and cold reservoir if the target system starts 
in a superposition of many energy eigenstates (as a coherent 
state in quantum optics). This should be considered as justification 
of our model.

In this paper, we do not consider the question of which
physical interactions could implement the desired unitaries.
It is clearly far away from present technology to implement
a unitary heat engine in such a way that the energy consumption of
the implementation is less than the thermodynamic energy yield.
The interesting question is whether this is a matter 
of principle or not;
if there are fundamental bounds on the energy consumption of
the required unitaries one should expect similar lower bounds
for logical operations. At the moment, no fundamental 
lower bounds
are known on the energy consumption of a computer
\cite{Benioff,Bennett:73,Feynman:85,
Margolus:90,viva2002,Ergodic}; likewise 
we do not know of any such bound for heat engines. 
Nevertheless  molecular heat engines 
extracting the {\it full amount} of
 thermodynamically available work from a
hot and a cold reservoir could  involve 
logical transformation too complex
to be feasible.

\section{Unitary Heat Engines}

In order 
to see that unitary processes on 
molecular systems with different temperature could extract some work
we first want to recall thermodynamics of quantum systems
with discrete energy levels. 
We will also give another characterization of thermal equilibrium states
as the only states which do not allow any energy extraction even if
arbitrarily many copies are available. 
The proof of this statement
shows that the
Boltzmann distribution in conventional thermodynamics
arises nicely 
from a geometric condition in the high dimensional state space
of many particle systems.
First we state the usual definition of thermal equilibrium
(``Gibbs state''):

\begin{definition}{\bf (Thermal Equilibrium)}\\
Let $H:=\sum_j E_j |j\rangle\langle j|$  be the Hamiltonian
with energy levels $E_j$ and energy eigenstates  $|j\rangle$. 
Whenever $tr(\exp(-H/T))$ is finite, 
the thermal equilibrium state $\gamma_T$ for some $T>0$ is given by
\[
\gamma_T :=e^{-H/T} /tr(e^{-H/T}) =\sum_j e^{-E_j/T}|j\rangle \langle j|/\sum_l e^{-E_l/T}\,,
\]
where we have dropped Boltzmann's constant. 
For $T=\infty$ the density matrix $\gamma_T$ 
is the maximally mixed state
and for $T=0$ a uniform mixture over all ground states, i.e., energy 
eigenstates with minimal energy.
\end{definition}

\noindent
One can check easily that $\gamma_T$ does not allow any 
energy extraction since the states with lower energy are more likely
than the states with larger energy.
Note that the converse statement is not true, i.e., there 
are states $\rho$ which differ from all temperature states
$\gamma_T$ for $T\in [0,\infty]$ but for which no unitary lowering the energy exists.
However, a weaker form of the converse is true:

\begin{theorem}
\label{Th:Haupt}{\bf (Copies of Non-Equilibrium States are Energy Sources)}\\
Let $\rho$ be a state with $\rho \neq \gamma_T$ for all
$T\in [0,\infty]$ which has the additional property that not
all the probability is concentrated in the ground states.
Then there is an $n\in \N$ such that an appropriate unitary 
extracts energy from $\rho^{\otimes n}$. 
\end{theorem}

\vspace{0.5cm}
\noindent
Proof:
It is clear that $U$ can only minimize 
$tr(U\rho U^\dagger H)$ if
$U\rho U^\dagger$ commutes with $H$  \cite{Bh}. 
We may therefore assume that
$\rho$ already commutes with $H$.
Let $p_0,\dots,p_{d-1}$ be the eigenvalues of $\rho$ corresponding
to the energy levels $E_0,\dots,E_{d-1}$. 
An eigenbasis of $\rho^{\otimes n}$ is clearly given by 
all products of $n$ eigenvectors of $H$.
We characterize these basis states 
by vectors $l\in \Z^k$ 
with
$l=(l_1,\dots,l_d)$ and $\tilde{l}=(\tilde{l}_1,\dots,\tilde{l}_d)$
where $l_j$ and $\tilde{l}_j$ are the number of components being 
in level $j$. Their energy difference can be written as an inner product
in $\R^n$:
\[
\sum_j (l_j-\tilde{l}_j) E_j =( l-\tilde{l}| E )\,.
\]
Assume first that $p_j\neq 0$ for all $j=1,\dots,d$. 
The logarithm of the probability ratio
of the two states
 can also be written as an inner 
product;
\[
\sum_j (l_j-\tilde{l}_j) \ln p_j =
(l-\tilde{l}|\ln p)\,,
\]
where `$\ln p$' denotes the vector obtained by taking the logarithm 
of each
entry of $p$. 
Assume that $p$ is not the equilibrium distribution. 
Then there exists, by definition, no $T>0, \mu\in \R$ such that
\[
p_j= e^{- E_j/T+\mu} \,\,\,\forall j\,,
\]
which would be equivalent to 
\[
\ln p= - \frac{1}{T} E +\mu a\,,
\]
if we define $a\in \R^d$ as the vector having only $1$ as entries.
Let $R$ be the projection onto the space $a^\perp$.
Then we have 
\[
R\,\,\ln p\neq - \frac{1}{T} R\,E 
\]
for all $T >0$.
Since not all levels have the same energy we have 
$R\,E\neq 0$. Furthermore, not all entries of $\ln p$ are equal
because  this would be the $T=\infty$ state. 
Elementary geometry shows that there is  an $x$ in $a^\perp$ such that
\[
( x| \ln p ) >0
\]
and
\[
( x| E ) >0\,.
\]
Of course $x$ can be chosen with rational entries and therefore, by multiplication with the least common multiple of their denominators,
also as a vector $x\in \Z^d$. 
With such an $x$ there exist vectors $l,\tilde{l} \in \N_0^d$ 
such that $x=l-\tilde{l}$. By defining 
\[
n:=\sum_j l_j\,,
\]
which is equal
to the sum of all $\tilde{l}_j$ according to $l-\tilde{l} \perp a$, 
the vectors
$l, \tilde{l}$ define two classes of  states  such that
each state in one class is more likely than each state in the other class
although the latter states have less energy.

Let $p$ have entries zero. Assume that there is some non-zero
probability for a level which is not the ground state.
Then there are 3 levels $0,1,2$ with
$E_2,E_1>E_0$ such that $p_2=0$, $p_1\neq 0$.  Consider
a state with label $l$ in $O^{\otimes n}$  
with $l_2=1,l_0=n-1$ and $l_j=0$ for all the other $j$.
Consider furthermore a state with label $\tilde{l}$ where
$\overline{l}_1=n$ and $\tilde{l}_j=0$ for all the other indices $j$.
 Clearly
there is an $n$ such that $\overline{l}$ has more energy than
$l$ even though the former has non-zero probability and the latter
probability zero. 
\hfill $\Box$

\vspace{0.5cm}

\noindent
In agreement with thermodynamic intuition, elementary calculation
shows that
the composition $\rho_A\otimes \rho_B$ of 
two equilibrium states 
$\rho_A,\rho_B$  with the same temperature is
the unique equilibrium state of the composed system.
This shows that
the $n$ fold copy of an equilibrium state still allows
no work extraction. 
We will clearly expect that if a state $\rho$ is  {\it close
to} an equilibrium state for some $T$ one will require
a {\it large number} of copies of $\rho$ to extract energy. 
We will later see that this fact implies that
two reservoirs consisting of hot and cold two-level systems,
respectively, require many-qubit operations whenever the temperature
difference between the two reservoirs is small.  
We now define the heat engine precisely:

\begin{definition}{\bf (Unitary Heat Engine)}
A  heat engine is a unitary transformation $U$ on a bipartite system
with Hamiltonian $H:=H_A \otimes {\bf 1} + {\bf 1} \otimes H_B$.
It  is initially 
in the state $\rho=\rho_A\otimes \rho_B$ where 
$\rho_A,\rho_B$ are equilibrium states with different temperatures
$T_A,T_B$, respectively and $U$ extracts energy in the sense
of Definition \ref{ExtrDef}.
A unitary  $U$ is an optimal heat engine if it maximizes
$E_{gain}$. 
\end{definition}

\section{SWAP as the Most Elementary Heat Engine}

\label{Elementary}

Classical thermodynamics states that one can in principle use
any two systems with different temperature to extract work 
in a Carnot cycle \cite{Callen}.
For two quantum systems this is no longer true if we
demand unitary heat engines as in Definition \ref{ExtrDef}.
If the systems $A$ and $B$ are two-level systems with equal
energy gap $E$ but different temperatures $T_A,T_B$ 
one checks easily that
the 4 states $00,01,10,11$ of the bipartite system
already satisfy  $p_{00}> p_{10},p_{01} > p_{11}$. Therefore
no work extraction is possible since this order 
coincides with ordering the states
according to their energy:
$E_{00} < E_{0,1}=E_{1,0}<E_{11}$.
However, we can construct a heat engine if 
the energy gaps $E_A$ and $E_B$ of system $A$ and $B$ 
satisfy
\begin{equation}\label{AddCon}
\frac{T_A}{E_A} > \frac{T_B}{E_B}\,.
\end{equation}
One observes easily that the state $10$ has more energy than $01$ even
though the latter is more likely due to
\[
\frac{p_{10}}{p_{01}}=e^{-E_A/T_A} e^{E_B/T_B}
=e^{(E_B/T_B - E_A/T_A)}\,.
\]
That the latter term should be greater than $1$ 
leads directly to Eq.~(\ref{AddCon}).
Then we can gain energy by implementation of the SWAP-gate, i.e.,
the permutation $10 \leftrightarrow 01$.
The condition (\ref{AddCon})  is
specific to our molecular heat engine
and seems not to be directly  related with the second law.
It is something like an additional constraint for microphysics;
such constraints become less
relevant in larger systems. 

It is easy to see that the SWAP gate is the only
possible unitary operation that extracts an {\it maximal} 
amount of energy since it is clear that the states $00$ and $11$ 
must remain unchanged as they already have both extremal energy and
probability.
Only for the two states $10$ and $01$ the 
order corresponding to increasing energy  
is not consistent with the order corresponding
to decreasing probability and we must exchange the states.

One could easily think of transformations $U$ which are close
to the unique optimal one. It is intuitively obvious that one could find
some trade-off relations between {\it efficiency} of the heat engine $U$ 
and its {\it reliability} as a SWAP gate.
This trade-off relation would also hold if convex sums of unitaries
(a ``random unitary heat engine'') were applied.

\section{Approximate Computation of Roots and Powers 
with Oscillator Modes}

To show that heat engines may involve quite complex transformations
we have to consider larger systems. 
A very natural system in physics is a quantum harmonic oscillator.
Its Hilbert space $l^2(\N_0)$ is spanned by the number states
$|0\rangle, |1\rangle, |2\rangle, \dots$ with $0,1,2,\dots$
quanta. Such a system can be a quantum optical  
mode or a mechanical oscillator.
A state with $j$ quanta of frequency $\omega$ has the
energy $E(j)=  j \hbar \omega $ and   the system Hamiltonian is
therefore
\[
H:=\hbar \omega  \sum_{j=0}^\infty j |j\rangle\langle j| \,. 
\]
The bipartite system on which our heat
engine will be defined consists of two modes with different
frequencies $\omega_A$ and $\omega_B$. In studying optimal heat engines
on such a 
 bipartite system a problem specific to infinite systems will arise:
We have usually constructed the optimal heat engine $U$ by 
forming two lists, one containing the basis states ordered
by decreasing probability and the other by increasing energy.
Then $U$ is given by the map $|a\rangle \mapsto |b\rangle$  
for each  corresponding pair $(a,b)$. 
Unfortunately the first list may be incomplete even though the other 
is complete. Then $U$ is  not defined
on all states. This situation occurs in the zero-temperature
case. Then for every bijection on the basis states there is always 
a bijection extracting more energy.
 Below we will ignore this problem because 
one can easily check that
the maps described there can be approximated
by unitary heat engines in an appropriate way.

First we assume $\omega_A=\omega_B=\omega$ and $T_A\neq T_B=0$.
Then the only states with non-vanishing probability are of the form
\[
|n,0\rangle\,\,\,\,\,\,\,n \in \N_0\,. 
\]
To construct the image of the state $|n,0\rangle$ 
with respect to an optimal heat engine 
we recall that the eigenspace of the joint Hamiltonian
\[
H\otimes {\bf 1} +{\bf 1} \otimes H
\]
is degenerate. Since $\omega$ is irrelevant in the following we
assume $\hbar \omega=1$ such that we obtain $\N_0$ as the spectrum.  
The eigenspace $\cH_k$ corresponding to eigenvalue $k\in \N_0$ has
 dimension $k+1$. The optimal heat engine $U$ has to map the state
$|n,0\rangle$ into an eigenspace $\cH_k$ where $k$ is uniquely specified
by the following conditions:
\[
dim (\oplus_{l<k} \cH_l) <n+1\,,
\]
and
\[
dim (\oplus_{l\leq k} \cH_l) \geq n+1\,.
\]
By calculating the dimensions we obtain
\[
n+1 > \sum_{l=0}^{k-1} (l+1)=\sum_{l=1}^k l =\frac{k^2+k}{2} \,,
\]
and
\[
n+1 \leq \sum_{l=0}^{k} (l+1)=\frac{k^2+3k +2}{2}\,.
\] 
The conditions are equivalent to
\[
(k^2+k)/2 <n+1 \leq (k^2+3k+2)/2\,.
\]
For large $n$ we have $k\approx \sqrt{n}$.
Hence we have a calculator which reduces the
approximative computation of $\sqrt{n}$ 
to an addition of  numbers
by the
following procedure: 
Apply the heat engine to the state $|n,0\rangle$ and measure
$\tilde{n},\tilde{m}$ of the resulting state.
Then $\tilde{n}+\tilde{m}$ is an estimation for $\sqrt{n}$. 

Two modes can also be used for the calculation of squares.
Let $\omega_B< c\omega_A$ with some real $c\gg 1$. Then 
choose the temperatures
such that $T_A= cT_B$. This implies that all states 
$|n,m\rangle,|n',m'\rangle$ with $n+m=n'+m'$ have equal probability.
The optimal heat engine on these two oscillators can compute 
approximately $(n+m)^2$ for the input
$n,m$ whenever  the result is sufficiently smaller than $c$.
This is seen as follows:

For a given pair $n,m$ with $n+m=k$ there
is a  $(k^2+k)/2$-dimensional space for which the eigenvalue
of the joint density matrix is greater than the eigenvalue 
corresponding to the state $|n,m\rangle$.  
This space must 
be mapped into the  span of all states $|0,\tilde{m}'\rangle$ with
$\tilde{m}' \leq (k^2+k)/2$. The subspace 
\[
\oplus_{l\leq k} \cH_l
\]
has to be mapped on the span of all   $|0,\tilde{m}'\rangle$ with
$\tilde{m}' \leq (k^2+3k+2)/2$.  
When we initialize the heat engine to a
state $|n,m\rangle$  and measure the right quantum number
we obtain therefore some $\tilde{m}$ with 
\[
\frac{k^2+3k+2}{2}     \geq  m \geq \frac{k^2+k}{2}\,.
\]
Hence we obtain $\tilde{m} \approx k^2/2$. 

The schemes above generalize in a straightforward 
way to the computation of 
higher powers and higher roots when more than two oscillators
are used. To calculate $k$th roots we start with $1$ hot and
$k-1$ zero-temperature oscillators and to calculate 
the $k$th power we start with $k$ hot and $1$ cold modes
where the hot modes have $c$ times larger energy gap and the 
temperatures
$T_A=c T_B$ are also chosen such that the probability for
a state $|n_1,n_2,\dots,n_k\rangle$ is only determined
by $N:=\sum_j n_j$. Then the optimal heat engine maps all states 
with $N<c$ onto a state $|0,0,\dots,\overline{N}\rangle$ where
$\overline{N}$ is an approximation for $\sqrt{k}{N}$.

\begin{figure}\label{KhodersBild}
\centerline{
\epsfbox[190 240 460 650]{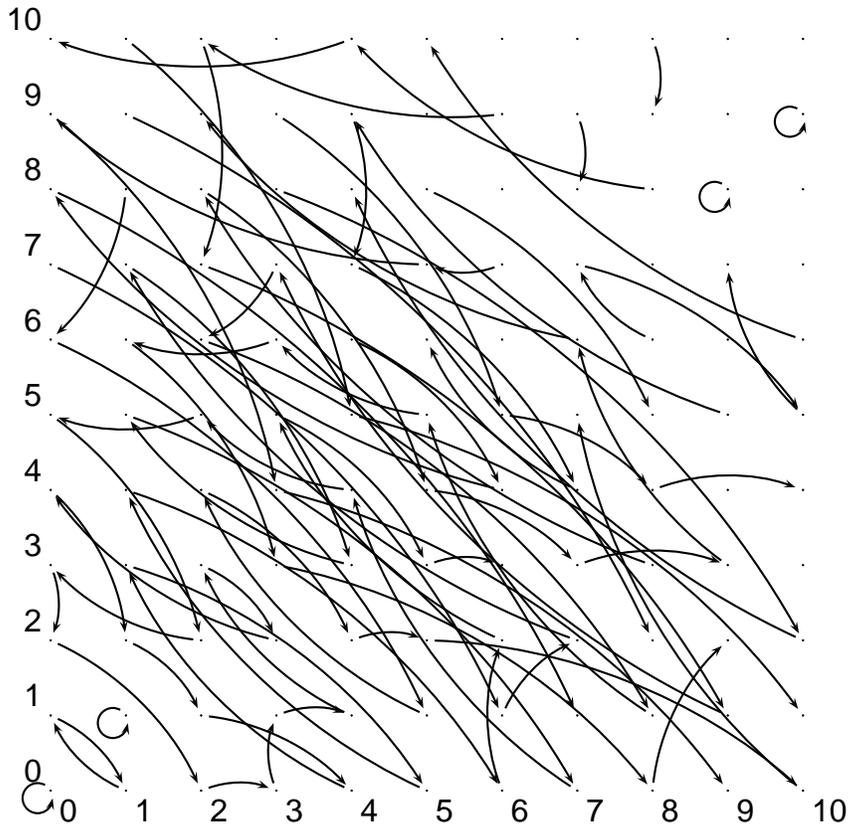}
}
\caption{{\small Optimal heat engine of two harmonic oscillators
with
frequency ratio $\omega_A/\omega_B\sqrt{2}$ and
 temperature
ratio $T_B/T_A=\sqrt{3} \omega_B/\omega_A$. An 
arrow $(n,m) \rightarrow (\tilde{n},\tilde{m})$ 
 indicates that a state with $n$ quanta in mode A and
$m$ in mode B has to be converted into a state
with $\tilde{n},\tilde{m}$ quanta, respectively. 
Points which have their
image or pre-image outside the depicted area obtain
no arrow.}}
\end{figure}

Now
 we assume that 
the ratio $\omega_A/\omega_B$ is irrational. This ensures that the 
Hamiltonian of the composite system is non-degenerate.
Up to irrelevant constants, 
the energy of a state with $n_A$ quanta in mode $A$ and 
$n_B$ quanta in mode $B$ is 
\[
E(n_A,n_B)= r n_A +n_B
\]
with $r\in \R \setminus \Q$. 
We define 
a bijective function $k:\N_0^2\rightarrow \N_0$ 
such that
$k(n_A,n_B)$ indicates the number of the pair $(n_A,n_B)$
when all pairs are put into an increasing order 
with respect to $E(n_A,n_B)$. 
Now we choose the temperatures $0\neq T_A\neq T_B\neq 0$ such that
\[
q:=\frac{E_A/T_A}{E_B/T_B}
\]
is also irrational which
holds for instance when $T_A/T_B$ is 
rational.  It follows that the density operator 
$\rho_A \otimes \rho_B$ is also non-degenerate.
Up to additive and multiplicative constants, 
the logarithm of the probability for a state 
$|n_A\rangle \otimes |n_B\rangle$ is given by 
\[
Q(n_A,n_B):=q n_A +n_B\,.
\] 
A larger value $Q(n_A,n_B)$ indicates that the state is less likely.
In analogy to the map $k$ we define a bijective 
function $l:\N_0^2\rightarrow N_0$ 
indicating the order 
of the pairs $(n_A,n_B)$  with respect to their values 
$Q(n_A,n_B)$. 
Define a permutation $\pi$  on $\N_0^2$ by
\[
\pi:=k \circ l^{-1}\,.
\]
This permutation of basis states $|n_A,n_B\rangle$ defines
a unitary $U_\pi$ by linear extension\footnote{Note that the ordering
of pairs given by $E$ or $Q$ is a term order in the sense of 
\cite{Weispfenning}}.
The density operator of the whole system 
after having implemented the heat engine $U_\pi$ is  
\[
U_\pi (\rho_A \otimes \rho_B) U_\pi^\dagger\,.
\]
The heat engine permutes 
the eigenvalues such that they are reordered according to 
the corresponding energy values. 
We have computed the corresponding reordering of states
for the values $q=\sqrt{2}$ and $e=1/\sqrt{3}$. The mapping is depicted in
Fig.~\ref{KhodersBild}, 
showing that the heat engine defines a quite {\it complex}
flow in the discrete two-dimensional plane.

Another interesting  case is when one temperature is zero. 
For $T_A=0$ only states $|0, n_B\rangle$ have non-zero 
probability. Here the number $n_B$ indicates already the 
ordering of all states which have non-zero probability. 
The optimal heat engine would have to map $(0,n_B)$ onto 
the state $(\tilde{n}_A,\tilde{n}_B)$ with $k(\tilde{n}_A,\tilde{n}_B)=n_B$. Hence the heat engine solves the computation problem
of {\it inverting} $k$.

\section{Converting between Different Number Systems
with
$N$-Level Systems}

The heat engines above could only be used for approximate
calculations. Here we present an example with some finite dimensional
systems which perform an exact computation. 
Consider 
two  $N_A$-level systems with temperature $T_A$ and two
$N_B$-level systems with temperature $T_B=0$. 
Let all $4$ systems have equidistant energy levels
where the first system of type $A$ has energy gaps $E_A$ and the second
$N_A E_A$. For the $N_B$-level systems  
we have energy gaps $E_B$ and $N_BE_B$ and  assume furthermore 
$N_B E_B <E_A$.  
Then the $N_B^2$ states with least energy are given by
\[
|0,0\rangle \otimes |\tilde{n},\tilde{m}\rangle \,,
\]
where the rightmost vector denotes the states of the 
two $N_B$-level systems. The energy is increasing according to
an increase of $\tilde{n} N_A+\tilde{m}$. 
The $N_A^2$ most likely states are given by
\[
|n,m\rangle \otimes |0,0\rangle\,,
\]
and their probability is decreasing with increasing $nN_A +m$.
It is easy to check that the optimal heat 
engine for this level spacing and this temperature configuration can 
convert natural numbers from the $N_A$-ary representation to the 
$N_B$-ary representation:
Initialize the system  to the state
\[
|n,m\rangle \otimes|0,0\rangle \,,
\]
such that $n N_A +m < N_B^2-1$. Then we obtain
a state 
\[
|0,0\rangle \otimes |\tilde{n},\tilde{m}\rangle
\]
such that $\tilde{n} N_B + \tilde{m} =n N_A +m$, i.e.,
the representation of the input number in the $N_B$-system. 
The scheme generalizes canonically to numbers with more than $2$ digits but
since the energy gaps grow exponentially this would not  be useful to
transform numbers with many digits. Nevertheless the example shows that
optimal heat engines could perform some useful calculations.

\section{Computer Scientist's Heat Engine}

\label{CS}

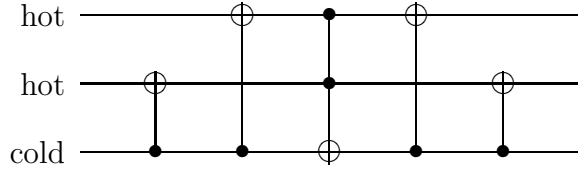
\begin{figure}
\centerline{
\Qcircuit @C=2em @R=1.4em {
&\lstick{\hbox{hot}} & \qw      &\targ   &\ctrl{2}& \targ &  \qw & \qw\\
&\lstick{\hbox{hot}}&  \targ    &\qw     &\ctrl{1}& \qw   &  \targ & \qw \\
&\lstick{\hbox{cold}}& \ctrl{-1} &\ctrl{-2}&    \targ& \ctrl{-2}& \ctrl{-1} &\qw
} 
}
\caption{{\small Heat engine with $3$ two-level systems with equal
energy gap.}}  
\label{JoesC}
\end{figure}

Note that
a system with two oscillator modes can never be an {\it efficient}
computer even though
it may perform some useful computations since the energy resource
requirements for representing an $n$ bit input increases exponentially
with $n$ instead of increasing only polynomially. 
Therefore $n$ qubits are more natural in order to study 
whether implementation of heat engines is close to computing.
The heat engine with two $2$-level systems studied 
in Section \ref{Elementary} requires
different energy gaps.
One can easily conclude from Theorem \ref{Th:Haupt} that heat engines
are also possible with two-level systems with equal energy gap
if one has a few of them:
The composition of $2$ two-level systems with  temperatures $T_A\neq T_B$
\[
\rho:= \gamma_{T_A} \otimes \gamma_{T_B}
\]
is not an equilibrium state for any temperature. Therefore,
there is an $n$ such that $\rho^{\otimes n}$ allows a unitary heat 
engine.
If the temperatures differ sufficiently this is already true
for $n=2$. One can even implement an heat engine 
with $2$ hot and $1$ cold system.
Assume $T_A > 2 T_B$.
One checks easily that the state $110$ is 
more likely than $001$ even though its
energy  is twice as much. Hence the process $110\leftrightarrow 001$ 
extracts some energy.
Fig.~\ref{JoesC} 
shows a quantum circuit implementing this heat engine.

The required number of systems which are necessary 
in order to make a heat engine possible at all
increases whenever the temperature quotient gets closer to $1$:

\begin{theorem}{\bf (Complexity of Using Small Temperature Gaps)}\\
\label{Th:Gebiet}
A heat engine on $n_A$ hot and $n_B$  cold qubits 
with temperatures $T_A$ and $T_B$, respectively, and equal energy gaps, 
is possible if and only if 

\begin{enumerate}

\item  (for $n_A \leq n_B$)
\[
\frac{T_A}{T_B} \geq \frac{n_A}{n_A-1}
\]

\item (for $n_A \geq n_B$)
\[
\frac{T_A}{T_B} \geq \frac{n_B+1}{n_B}
\]   

\end{enumerate}
Furthermore, every heat engine acting on an infinite reservoir of
hot and cold qubit level systems must use operations which connect
at least $n_A$ hot and $n_B$ cold systems such that the above conditions
hold. 
\end{theorem}

\noindent
{\bf Proof:}
We note that a heat engine can  work 
if and only if a pair of states exist such that the first has more energy
even though it  is more likely.
Let $(l_A,l_B)$ denote the Hamming weights of a basis state in the 
$n_A+n_B$ qubit system.  The pair $(l_A,l_B)$ and $(k_A,k_B)$ 
satisfies this condition if 
\[
(l_A-k_A) -(l_B-k_B) >0
\]
and
\[
(l_A-k_A)T_A -(l_B-k_B)T_B <0
\] 
Elementary computation shows that this implies
\[
\frac{T_A}{T_B} > \frac{l_A-k_A}{l_B-k_B} > 1\,.
\]
Clearly the modulus of the numerator and the denominator
are at most  $n_A$ and $n_B$, respectively. 
The smallest possible quotient which is still greater than  $1$ is therefore
$n_A/(n_A-1)$ or $(n_B+1)/n_B$, respectively.
This shows that the conditions (1), respectively (2) are 
necessary in order to make
a heat engine possible.

For the converse we observe that in case (1)  a permutation of the states
$(n_A,0)$ and $(0,n_A-1)$ extracts some amount of 
energy. In case (2) one extracts
energy by permuting 
$(n_B+1,0)$ and $(0,n_B)$.
$\Box$.

\begin{figure}
\label{fig:Gebiete}
\centerline{
\epsfbox[0 0 71 185]{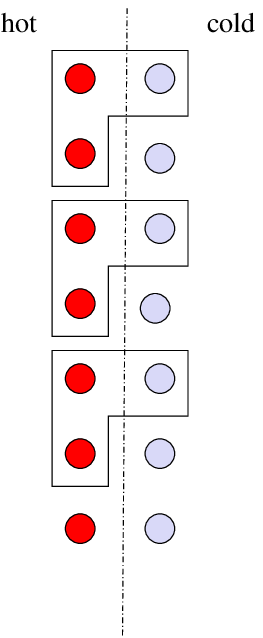} \hspace{3cm}
\epsfbox[0 0 71 185]{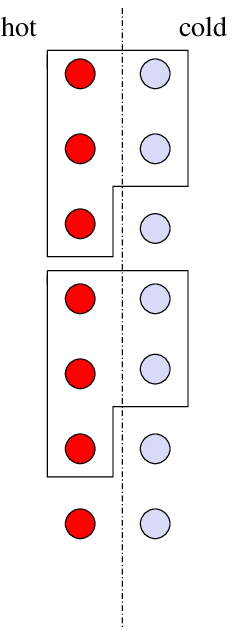}\hspace{3cm}
\epsfbox[0 0 71 185]{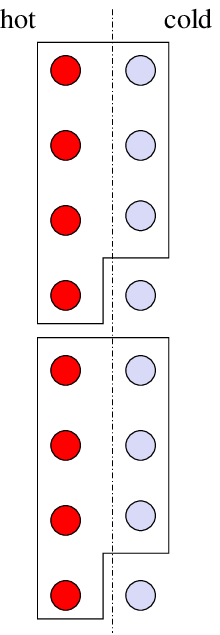}
}
\caption{{\small Heat engines with  $T_A/T_B >2$ 
 can be implemented with joint operations on $2$ hot 
and $1$ cold qubit (left). For $2>T_A/T_B > 3/2$ operations
on $3$ hot and $2$ cold qubits are needed (middle), and heat engines 
for $3/2 > T_A/T_B >4/3$ must involve 
$4$ hot and $3$ cold qubits (right).}}
\end{figure}

\vspace{0.5cm}
\noindent
Fig.~\ref{fig:Gebiete} 
illustrates how the complexity of heat engines
on two-level systems with equal energy gap 
 increases when the temperature gaps decrease
in the sense that more qubits have to be involved.
Note that Fig.~\ref{fig:Gebiete} furthermore sketches a
 simple method to obtain
{\it suboptimal} heat engines on many particles by independently 
applying few-qubit
heat engines.

\begin{figure}
\label{3Circuit}
\centerline{
\Qcircuit @C=2em @R=1.4em {
&\lstick{\hbox{hot}} & \qw & \ctrl{1} &  \targ   & \qw     & \qw \\
&\lstick{\hbox{hot}} & \qw & \ctrl{2} &  \ctrl{-1} & \targ  & \qw  \\
&\lstick{\hbox{hot}} & \qw & \qw     &  \qw      & \ctrl{-1}& \qw \\
&\lstick{\hbox{cold}} & \qw & \targ  & \ctrl{-3} &   \ctrl{-2}& \qw
} \hspace{2cm}
\Qcircuit @C=2em @R=1.4em {
&\lstick{\hbox{hot}} & \qw & \ctrl{1} &  \ctrl{1}  & \targ      & \ctrl{1}& \qw \\
&\lstick{\hbox{hot}} & \qw & \ctrl{2} &  \targ  &   \qw     &
\ctrl{2} & \qw \\
&\lstick{\hbox{hot}} & \qw & \qw     &   \qw & \ctrl{-2}    & \qw & \qw
 \\
&\lstick{\hbox{cold}} & \qw & \targ  &   \ctrl{-2}&  \ctrl{-3} & \targ &
\qw
}
}
\caption{{\small (left:) The simplest possible heat engine which uses 
only
Toffoli gates.  
(right:) A heat engine with $4$ gates can extract work
from reservoirs with smaller temperature gaps.}}
\end{figure}
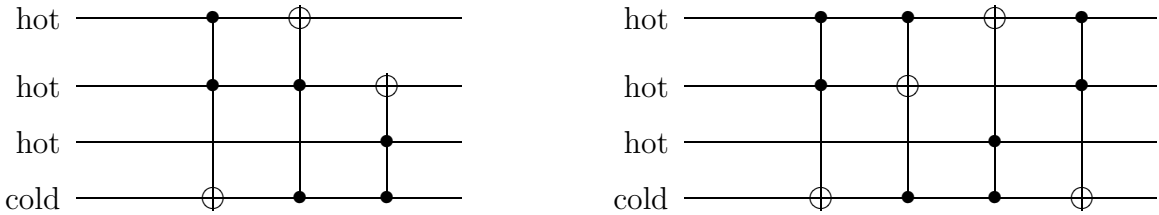

We would like to know how this increase of complexity
with decreasing temperature difference 
occurs also with respect to the
{\it number of required gates} in a circuit which consists
of simple elementary gates. Therefore we have checked 
which heat engines are possible when only a few number gates 
are allowed and restricted the attention to TOFFOLI gates 
\cite{Toffoli}
since (1) they permute only basis states and can therefore 
easily be treated with computer algebra systems
(2)  they are universal for classical
computation and they seem therefore sufficiently powerful 
to generate good heat engines\footnote{Note, however, that TOFOLLI gates
do not generate the full group $S_N$ of permutations on $N=2^n$ basis states of $n$ qubits (see Section \ref{Majority}).}. 
The restriction to one type of gate 
simplifies the complete search for all circuits which can be obtained
with $k$ gates. 
In the first experiment we consider $3$ hot and $2$ 
qubits with temperatures $T_A=E /\ln 2$ and $T_B=0$. 
The probabilities for the upper state is hence 
for a hot qubit given by $1/3$.
Computer algebra calculations have shown that $3$ 
gates are necessary in order to have positive energy gain.
The circuit Fig.~\ref{3Circuit}, left, acts on only $3+1$ qubits
and does not make use of the second available cold qubit.
With this circuit we have
$E_{gain}=1/27$ which is
considerably less than  the 
optimal heat engine on $3+2$ qubits having $E_{gain}=5/27$.
 If we increase the temperature of the cold system
such that $T_B= E /\ln 5$ the circuits with $3$ gates 
do not decrease the average energy any more and at least
$4$ gates are needed for a heat engine. 
One 
possibility with $4$ gates is shown in Fig.~\ref{3Circuit}, right, with 
$E_{gain}=1/72$.
For temperature $T_B = E  /\ln 4$ there is even no heat engine
with $5$ Toffoli gates. Note that 
a heat engine is possible  by Theorem \ref{Th:Gebiet} because
$T_A/T_B=\ln 4 /\ln 2=2 >3/2$; therefore the state exchange
$|111\rangle \otimes |00\rangle \leftrightarrow 
|000\rangle \otimes |11\rangle$ extracts energy.
Computer algebra calculations show furthermore  
that there is indeed
a circuit with 31 Toffoli gates  which  implements
the state exchange above such that all the other states are mapped 
onto basis states with the same Hamming weight.  
There are probably much simpler circuits but we know that
at least $6$ are required.

\section{MAJORITY Gate on 2n+1 Qubits} 

\label{Majority}

Here we present an example where the complexity 
of the optimal heat engine
can be compared to the complexity of a circuit which computes a 
well-known boolean function. 
Consider $2n$ two-level systems with temperature $T_A=\infty$ and $1$ system
with $T_B=0$. The basis states of the system are binary words
of length $2n+1$. The joint Hamiltonian of the system
is given by
\[
H:=E\sum_b wgt(b) |b\rangle \langle b|\,,
\]
where $E$ is the energy gap of each two-level system 
and $wgt(b)$ denotes the Hamming weight of the binary word $b$. 
Let the suffix of each of this binary words indicate the
state of system $B$. 
Then all binary words with suffix $0$ have
probability $1/2^n$ and words with suffix $1$ never occur.
Every optimal heat engine $U$  has to map the subspace spanned by the
former $2^n$ words onto the subspace corresponding
to the $2^n$ smallest  eigenvalues of $H$. It is the space spanned by
all words with Hamming weight at most $n$.
Therefore the inverse of the heat engine, i.e., $U^{-1}$ 
computes the boolean function MAJORITY in the sense that 
the rightmost qubit in the state
\[
U^{-1}|b\rangle
\]
is $1$ if and only if $wgt(b)>n$, i.e., the majority of
the qubits are in the $1$ state.
We would like to estimate the gate complexity of $U$
when it is implemented by elementary gates. If the set of elementary gates contains with every gate also its inverse the complexity of 
$U$ and $U^{-1}$ coincide.
To obtain a lower bound on the circuit complexity
we could therefore use bounds on the circuit complexity 
of MAJORITY. In \cite{Hastad} one can find 
bounds for classical circuits
with bounded depth which consist of AND and OR with arbitrary fan-in.
We can give a lower bound on the circuit depth 
which holds for arbitrary $k$-qubit gates.
The observable which measures whether the suffix of a binary word
is $1$ or $0$ is $A:={\bf 1}_{2n}\otimes \sigma_z$.
It is obviously a $1$-qubit observable. The observable 
$U A U^\dagger$ 
which measures whether
the majority of qubits are $1$ is a $2n+1$-qubit observable.
In \cite{JB00c} we have argued that a circuit of depth $l$ can convert
a $1$-qubit observable at most into a  $k^l$-qubit observable.
Therefore we otain
\[
l\geq log_k (2n+1)
\]
as lower bound on the depth. This shows after all that 
the depth must necessarily increase with $n$ even though logarithmic growth
would be quite slow. We summarize:

\begin{theorem}{\bf (Lower Bound on the Depth)}\\
Let $U$ be an optimal heat engine on $2n$ two-level systems with
temperature $T_A\neq 0$ and one two-level system with $T_B=0$
where all $2n+1$ systems have the same energy gap. Then 
the implementation of $U$  with $k$-qubit gates requires at least
a circuit of depth $\log_k(2n+1)$.
\end{theorem}

\noindent
Note that no ciruit which consists only of NOT and CNOT can implement $U$.
The action of CNOT permutes the basis states of two qubits according
to 
\[
|a_1\rangle\otimes |a_2\rangle  
\mapsto |a_1\rangle \otimes |a_1\oplus a_2\rangle\,.
\]
If one identifies the pair $(a_1,a_2)$ with a vector
in a two-dimensional space over $F_2$ 
CNOT is a 
$F_2$-linear map. 
By embedding the action of CNOT into a $2n+1$-dimensional space 
over $F_2$ it remains linear. The action of NOT on qubit $j$ 
corresponds to
adding the vector $0\dots 010\dots 0$ with $1$ at position $j$.
Therefore every circuit $V$  with CNOT and NOT gates acts as
\[
V|b\rangle =|Ab +c\rangle\,,
\]
where $A$ is a $2n+1\times 2n+1$-matrix over $F_2$ and 
$c$ a vector in $F_2^{2n+1}$. 
If the majority function would be affine 
it could up to an additive constant 
be written as an inner product over $F_2$, i.e., 
there 
existed a vector $v$ and a number $w\in F_2$ 
such that $( v|b) \oplus w$
is $1$ whenever $wgt(b)>n$.
This is certainly not the case.
It is also easy to see that TOFFOLI gates 
alone cannot be sufficient to implement $U$ or $U^{-1}$.
Otherwise $U$ and $U^{-1}$ would leave all binary words with Hamming weight
at most $1$ invariant. But the state $|0\dots 01\rangle$ has to be mapped into
the space spanned by words with Hamming weight greater than $n$.

The insight that the inverse 
 of $U$ would  compute MAJORITY gave some 
hints on its complexity, however
it does not show that the heat engine itself can be used for computing
this boolean function. 
A thermodynamic machine which can directly be used
as a MAJORITY gate is the reverse of the heat engine, namely a
refrigerator. Assume we have given $2n+1$ two-level systems
with the same temperature $T\neq 0,\infty$. Then an optimal refrigerator
for the rightmost qubit is a transformation $U$ which reduces
the probability for its upper state as much as possible. 
This is certainly the case only when $U$ maps all states $|b\rangle$ with
$wgt(b)>n$ to the subspace $S$  spanned by words with suffix $1$ and
all with $wgt(b)\leq n$ to the orthogonal complement of $S$. 
Hence the rightmost qubit is the output qubit of a MAJORITY
computation. 
We summarize:

\begin{theorem}{\bf (Relation to Complexity of MAJORITY)}\\
Let $U$ be an optimal heat engine on $2n$ two-level systems with
temperature $T_A\neq 0$ and one two-level system with $T_B=0$
where all $2n+1$ systems have the same energy gap. Then 
the implementation of $U$ requires
at least as many elementary quantum gates as a computation of
the function MAJORITY of the $2n+1$-qubit input requires
which uses
no additional memory space. 
\end{theorem}

\section{Universal Classical Computation on Pairs 
of $3$-Level Systems}

It is clear that every optimal heat engine on $2$ two-level
systems leaves the states $00$ and $11$ invariant because
they are the states with minimal and maximal energy
and with maximal and minimal probability, respectively, 
at the same time. Therefore the only non-trivial
logical operation is a SWAP-gate. To find more interesting
logical gates in a heat engine on a bipartite system
we will therefore consider two $3$-level systems $A$ and $B$.
We assume that system $A$ and $B$ have both 
equidistant levels $|0\rangle,|1\rangle,|2\rangle$    
with energy gaps $E_A$ and $E_B$, respectively. 
Up to an irrelevant factor the  energy of a
state $|n,m\rangle$ with $n,m=\{0,1,2\}$ is given by
\[
E(n,m)=e n +m
\]
with $e:=E_A/E_B$.
The inverse logarithm 
of the probabilities is, up to irrelevant
additive and multiplicative constants, given by
\[
Q(n,m)=q n+m
\]
with $q:=E_A T_B /(E_B T_A)$.
When $e$ and $q$ are not in $\{1/2,1,2\}$ 
the Hamiltonian as well as the 
density matrix of the bipartite system 
are non-degenerate and the optimal heat engine implements 
a unique reordering of basis states.
The following choice of values $e,q$ turns out to be 
useful: setting $1<e<2$ we induce an order on energy values
of the pairs $n,m$  
which is a refinement of the degenerate order induced
by $n+m$ such that for pairs with equal $n+m$ 
preference is given to the pair with smaller $m$.
Explicitly, this is the order
$00,10,01,20,11,02,21,12,22$.
With $q>2$ the probabilities are in the lexicographic order
$00,01,02,10,11,12,20,21,22$.
 By comparing these
orders one checks easily that the optimal heat engine 
 implements the map
\begin{eqnarray}
\label{Map}
00 &\mapsto& 00 \\ 
01 &\mapsto& 10 \nonumber \\ 
02 &\mapsto& 01 \nonumber\\ 
10 &\mapsto& 20 \nonumber\\ 
11 &\mapsto& 11 \nonumber\\ 
12 &\mapsto& 02 \nonumber\\ 
20 &\mapsto& 21 \nonumber\\ 
21 &\mapsto& 12 \nonumber\\ 
22 &\mapsto& 22  \nonumber\,.
\end{eqnarray}

We will show that
the ability to implement this heat engine   on 
every  pair of 3-level systems
consisting of one system of type $A$ and one of type $B$
implies the ability to implement classical computation
on the collection of these $3$-level systems. 
For doing so, we chose
the encoding such that the logical states $0$, $1$ are
the states $|1\rangle$ and $|2\rangle$, respectively and
obtain a universal set of logical operations as follows:

\begin{enumerate}
\item {\bf OR from A, B to B:}

\centerline{
\epsfbox[0 0 140 66]{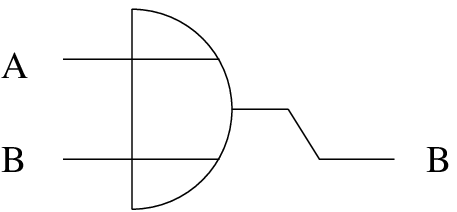}
}

\noindent
Apply $U$ once. One checks easily on tabular (\ref{Map}) that the second 
state is $|2\rangle$ if the input is one of the states
$|12\rangle, |21\rangle, |22\rangle$  and $|1\rangle$ if
the input is $|11\rangle$.

\item {\bf WIRE from
A to B:}

\centerline{
\epsfbox[0 0 140 55]{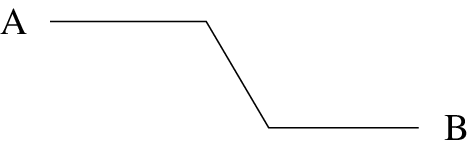}
}

\noindent
Use our OR gate  by initializing B to $|1\rangle$, i.e., the logical
$0$ state.

\item{\bf FANOUT from B to A, B:}

\centerline{
\epsfbox[0 0 140 55]{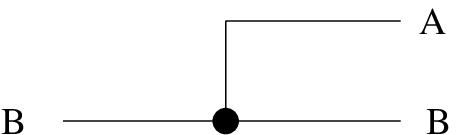}
}

\noindent
Initialize system $B$ to $|1\rangle$. Apply $U$ 4 times.
We get the mapping $12 \mapsto 20$ and  $11 \mapsto 11$.
The output on $A$ coincides already with the input on $B$.
The output on $B$ is $1$ or $0$ according to whether the input 
 on $B$ was $1$ or $2$. Hence the information has already been
copied to $B$ but with the wrong encoding.
For the decoding we
initialize an additional system $A'$ to the state $|1\rangle$ 
and apply $U^4$ to $A', B$.  We get $10 \mapsto 02$ and
$11 \mapsto 11$. Hence $B$ agrees with the original input on $B$.

\item {\bf WIRE from B to A:}

\centerline{
\epsfbox[0 0 140 55]{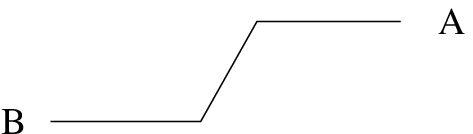}
}

\noindent
Use the FANOUT.

\item{\bf NOT from B to B:}

\centerline{
\epsfbox[0 0 140 40]{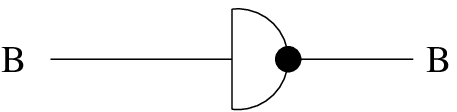}
}

\noindent
Implement the first part of the FANOUT operation which 
changes the input state $2$ to $0$ and leaves
$1$ invariant. By Initializing an additional system $A'$ 
to $|2\rangle$ and apply $U$ once we decode and negate
the information on $B$ simultaneously: $20\mapsto 21$ and
$21 \mapsto 12$.

\end{enumerate} 

\noindent
These operations allow obviously universal computation since 
every boolean function can be computed from circuits which consist only
of NOR gates.
We summarize:

\begin{theorem}{\bf (Universal Computing with Heat Engines
on  3-level Systems)}\\
Given two reservoirs  of  3-level systems
with temperature $T_A$ and $T_B$ and  energy gap $E_A$ and $E_B$, 
respectively, such that
\[
2\frac{T_A}{E_A} <\frac{T_B}{E_B}
\]
and
\[
2 >\frac{E_A}{E_B} > 1\,,
\]
then the ability to implement the optimal heat engine
on any chosen  pair  which consists of one system of type A 
and one of type B implies the ability to implement
universal classical computation on the 3-level system.
\end{theorem}

\section{Optimal Heat Engine as NP-Solver}

So far we have only shown some examples of heat engines which 
{\it perform} relatively simple logical operations or computations
or heat engines which are {\it generated by} some simple operations. 
Now we will consider an instance of a heat engine which
had to solve a computationally hard  problem in order to be optimal.
Consider 
a collection of two-level systems with different
energy gaps where one them has temperature $T_A$ and all the others
have temperature $T_B$. Then the optimal heat engine
is a computer which 
solves an NP-complete problem:

\begin{theorem}
Let $E_A,E_1,E_2,\dots,E_n$ be the energy gaps of $n+1$ two-level systems.
Let $T_A$ be the temperature of the $0$th system and $T$ of the 
remaining $n$. Let the values be such that 
there is no $b\in \{0,1\}^n$ such that
\[
\sum_{j=1}^n b_j E_j = \frac{E_A T_B}{T_A}\,.
\]
Let $U$ acting on $\C^2\otimes (\C^2)^{\otimes n}$ be an optimal unitary heat engine
for this system. Then $U$ solves a  KNAPSACK problem
in the following sense. Measure the right $n$ two-level systems 
with the state
\[
U(|1 \rangle \otimes 0 \dots 0\rangle) \,,
\]
in the computational basis.
Let $b\in \{0,1\}^n$ be the obtained result.
Then $b$ satisfies
\begin{equation}\label{Ungl}
E_A  > \sum_j b_j E_j >E_A\frac{T_B}{T_A}
\end{equation}
if and only if such a binary word $b$  exists.
\end{theorem}

\noindent
{\it Proof:} 
We only have to show that if there is a string $b$ satisfying 
(\ref{Ungl})  it will always show up as a measured result.
Write $( b| E)$ for 
$\sum_{j=1}^n b_j E_j$.
Intuitively, inequality (\ref{Ungl}) 
implies that the state $|0\rangle \otimes |b\rangle$ 
has less energy than $|1\rangle \otimes |0\rangle$ 
but is less likely.
The assumption 
 that 
there is no $b\in \{0,1\}^n$ such that
\[
( b|E ) =\frac{ E_A T_B}{T_A}\,
\]
means that $|1\rangle \otimes |0\rangle$ is an
eigenvector in a non-degenerate subspace. This uniqueness of
the eigenvalue ensures that 
the optimal heat engine does not have `too much choice' on which states
the state $|1\rangle \otimes |0\rangle$ has to be mapped to.

First consider the subspace $\cH_1$ 
spanned by all vectors $|0\rangle \otimes |c\rangle$ with $(E|c) < E_A T_B/T_A$.
The restriction of $\rho$ 
to this subspace (note that this is indeed an invariant subspace of $\rho$)
is left invariant by every optimal $U$ since the ordering of eigenvalues
of $\rho$ and of $H$ coincide here. All these states are more likely
than $|1\rangle \otimes |0\rangle$ and have less energy.

Binary words $c$ with $( E|c ) =E_A T_B/T_A$ do not exist
by assumption. 
 Now 
consider the subspace $\cH_2$ 
spanned by all  $|0\rangle \otimes |c\rangle$ with 
$ E_A>  \langle E|c\rangle  >E_A T_B/T_A$. They have less energy than
$|1\rangle \otimes |0\rangle$ but they are also less likely, i.e.,
all eigenvalues of $\rho$ on this subspace are smaller than
the eigenvalue $p$ of $\rho$ for the eigenvector 
$|1\rangle \otimes |0\rangle$. Note that $\cH_1$ and $\cH_2$ are spectral
subspaces of the total Hamiltonian, i.e., the only states with 
energy in the specified intervals are states $|0\rangle \otimes |c\rangle$ 
with $c$ satisfying the considered inequalities.
Hence every optimal $U$ has to 
map $|1\rangle \otimes |0\rangle$ into $\cH_2$ since its eigenvalue $p$ 
is
the largest one  except from the eigenvalues which have already filled
the lower spectral subspace $\cH_1$. \hfill $\Box$ 

\vspace{0.3cm}

\noindent
Clearly it is essential for the proof that the heat engine is optimal.
Efficient algorithms for suboptimal heat engines are  possible.
Note that it is is a well-known phenomenon in the theory of NP-complete
optimization problems that a slightly relaxed demand on the  optimality
allows already efficient approximations \cite{Ausiello}. 
To implement a suboptimal heat engine one can implement 
 the few-qubit
heat engines considered in Section \ref{CS} involving only 
some of the qubits in the reservoir $B$.

\section{Including the Target System}

\label{Just}

So far we have considered systems which are informationally 
closed in the sense that only unitary operations are available
but which are at the same time not energetically closed
since it was exactly our goal to extract energy from the system.
Our above justification for such a model was that we do not want to
allow  the energy sink to {\it absorb entropy} because this could
trivialize the whole problem. The following  paradox arises from
this justification: The 
energy extraction in the above
unitary heat engines is only a probabilistic phenomenon
since they decrease only the {\it average} energy of the system.
For some  energy eigenstates of the system are mapped onto states
with higher energy and some onto lower energy states. 
This implies that the energy of the target system is decreased
or increased, depending on the state of the system. 
Such a probabilistic change of the energy of the target
increases necessarily its entropy even though this was exactly
what we wanted to avoid.  Now we show that there are natural situations
where this entropy increase  is negligible. 
Let for simplicity the eigenvalues of the system Hamiltonian
$H_s$  be 
some integers and the energy spectrum of the
target be $\Z$, i.e.,
its Hamiltonian $H_t$ on $l^2(\Z)$ be given by
\[
H_t |j\rangle = j|j\rangle 
\]
for all $j\in \Z$. 
 Let $U$ be some unitary heat engine 
permuting energy eigenstates of $H_s$ and $(P_j)$ be a complete 
set of orthogonal projections on the energy eigenstates of $H_s$. 
Let $\Delta(j)$ be the energy difference between the eigenvalues
corresponding to $P_j$ and to $U P_j U^\dagger$
and $S^k$ be the shift of $l^2(\Z)$ defined by 
\[
S^k|j\rangle =|k+j\rangle\,.
\]
Then we define a unitary operation $V$ on the
joint Hilbert space 
\[
\cH_s \otimes l^2(\Z)\,,
\]
of $s$ and $t$  by
\[
V:= (U\otimes {\bf 1}) \sum_j P_j \otimes S^{\Delta(j)}\,.
\]
One checks easily that $V$ commutes with 
the 
joint Hamiltonian
\[
H:=H_s\otimes {\bf 1} +{\bf 1} \otimes H_t\,.
\]
Furthermore, we choose an initial state 
 vector 
$|\psi\rangle \in l^2(\Z)$ of the energy sink 
such that $\langle \psi | S^j \psi\rangle \approx 1$ for all 
those $j$ which occur as energy difference between initial and final
state of the system. 
Then 
the completely positive positive map $G$  given by
\[
G(\rho):=tr_M( V(\rho\otimes |\psi\rangle \langle \psi|)V^\dagger) 
\]
coincides almost with the unitary operation 
$\rho \mapsto U\rho U^\dagger$. The reason is that
a superposition of all the  eigenstates in a large interval of energy
values is insensitive to energy increase or decrease and
obtains therefore almost no information about the energy gain.  

This shows that the restriction of the energy conserving unitary $V$ 
to $s$ 
can indeed approximately  be the unitary $U$,
when the initial state of the energy sink is a pure state
with large energy spread.  If the Hilbert space of the
energy sink is replace by $l^2(\N_0)$ the construction of unitaries
and initial vectors which yield, by restriction, approximately 
the unitary $U$ is technically a bit more difficult. 
A coherent state in quantum optics 
which has large photon number expectation 
would be a physical example for a state
with large energy spread.
We conclude that
the unitary heat engine appears as a limit with macroscopic control 
fields and is therefore a consistent model. 
 
The statement that $V$ does not increase the entropy of $t$ 
by an considerable amount holds also when the initial state
of $t$
is a mixture of energy eigenstates over a large interval 
of energy values. One checks easily that the restriction of $V$
to $s$ is no longer close to a unitary operation. Instead, it
destroys superpositions between all those energy eigenstates
with different $\Delta (j)$. Nevertheless it permutes
the basis states in the same way as the unitary heat engine $U$ 
does which implies that it implements 
the same classical computation steps
as the unitary model would do.

\section{Conclusions}

Using several examples of toy heat engines we have shown that there
is, on the molecular scale, 
a strong coincidence  
between the task of  {\it computation}
and the task of {\it energy extraction} from heat reservoirs. 
The ability to extract
a {\it maximal} amount of work even requires 
operations for some systems even 
 which solve hard
computational problems.
Even though
suboptimal heat engines may in general 
not require computationally hard 
operations we have argued that work extraction from two-level 
systems 
with {\it almost} equal 
temperatures require {\it many-qubit} operations. 
We conclude that heat engines which extract work from
reservoirs with similar temperature 
require relative complex physical processes.

Clearly, we do not expect that future heat engines on the molecular
scale will be implemented by the type of gates we have considered.
However, the `Strong 
Church Turing Thesis' \cite{Vergis,Shor} states
that {\it any physical device can be simulated by a Turing machine 
in a number of steps polynomial in the resources used by the
computing device}. The quantum version of this
replaces the classical Turing machine with a {\it quantum} Turing  
machine \cite{Shor}. Believing in this principle, one should expect that
every process implementing a heat engine which solves an NP-complete problem
has an efficient simulation on a quantum computer. 
Provided that one does not believe in efficient quantum algorithms
for NP-hard problems, our  
results indicate therefore that there are complexity-theoretic
limitations to the efficiency of heat engines on the molecular scale.

\section*{Acknowledgments}

Thanks to Anja Groch for checking the possible of 
`heat engine circuits'
with computer algebra and also to Markus Grassl for some of the
calculations and helpful remarks. 
Khoder El-Zein computed the optimal heat engines
on two 3-level systems and on 2 oscillator modes. Interesting
discussions with Joe Renes are also acknowledged; he
found the circuit in Fig.~\ref{JoesC}.
Rainer Steinwandt has pointed out the connection between our ordering
of two-mode states and well-known
questions of term ordering. 

This work has been funded by the BMBF project 01/BB01/B.

\end{document}